\documentclass[12pt]{article}

\newenvironment{wileykeywords}{\textsf{Keywords:}\hspace{\stretch{1}}}{\hspace{\stretch{1}}\rule{1ex}{1ex}}

\usepackage{amsmath,amssymb}
\usepackage{graphicx}
\usepackage{color}
\usepackage{dcolumn}
\usepackage{bm}
\usepackage[numbers,super,comma,sort&compress]{natbib}

\usepackage{xcolor}
\usepackage[section]{placeins}
\usepackage{longtable}
\usepackage{color}
\usepackage{hyperref}

\usepackage{xparse}
\ExplSyntaxOn
\NewDocumentCommand{\longdash}{ O{2} }
 {
  --\prg_replicate:nn { #1 - 1 } { \negthinspace -- }
 }
\ExplSyntaxOff

\newcommand*\samethanks[1][\value{footnote}]{\footnotemark[#1]}

\definecolor{background-color}{gray}{0.98}

\title{\texttt{PyCDFT}: A Python package for constrained density functional theory}
\date{}

\author{He Ma\thanks{Department of Chemistry, University of Chicago, Chicago, Illinois 60637, United States} \thanks{Materials Science Division and Center for Molecular Engineering, Argonne National Laboratory, Lemont, Illinois 60439, USA}, Wennie Wang\thanks{Pritzker School of Molecular Engineering, University of Chicago, Chicago, Illinois 60637, United States}, Siyoung Kim\samethanks[3], Man-Hin Cheng\samethanks[3]~\thanks{Current Address: Department of Physics, ETH Zurich, 8093 Zurich, Switzerland}, \\ Marco Govoni\samethanks[2]~\samethanks[3], Giulia Galli\samethanks[1]~\samethanks[2]~\samethanks[3]
}

\begin{document}

\maketitle

\begin{abstract}
We present \texttt{PyCDFT}, a Python package to compute  diabatic states using constrained density functional theory (CDFT).
\texttt{PyCDFT} provides an object-oriented, customizable implementation of CDFT, and allows for both single-point self-consistent-field calculations and geometry optimizations.
\texttt{PyCDFT} is designed to interface with existing density functional theory (DFT) codes to perform CDFT calculations where constraint potentials are added to the Kohn-Sham Hamiltonian.
Here we demonstrate the use of \texttt{PyCDFT} by performing calculations with a massively parallel first-principles molecular dynamics code, \texttt{Qbox}, and
we benchmark its accuracy  by computing the electronic coupling between diabatic states for a set of organic molecules.
We show that \texttt{PyCDFT} yields  results in agreement with existing implementations and is a robust and flexible package for performing CDFT calculations.
The program is available at \url{https://github.com/hema-ted/pycdft/}.
\end{abstract}

\begin{wileykeywords}
constrained density functional theory, charge transfer, electronic coupling, diabatic states, Python
\end{wileykeywords}

\clearpage

\begin{figure}[h]
\centering
\colorbox{background-color}{
\fbox{
\begin{minipage}{1.0\textwidth}
\includegraphics[width=70mm,height=50mm]{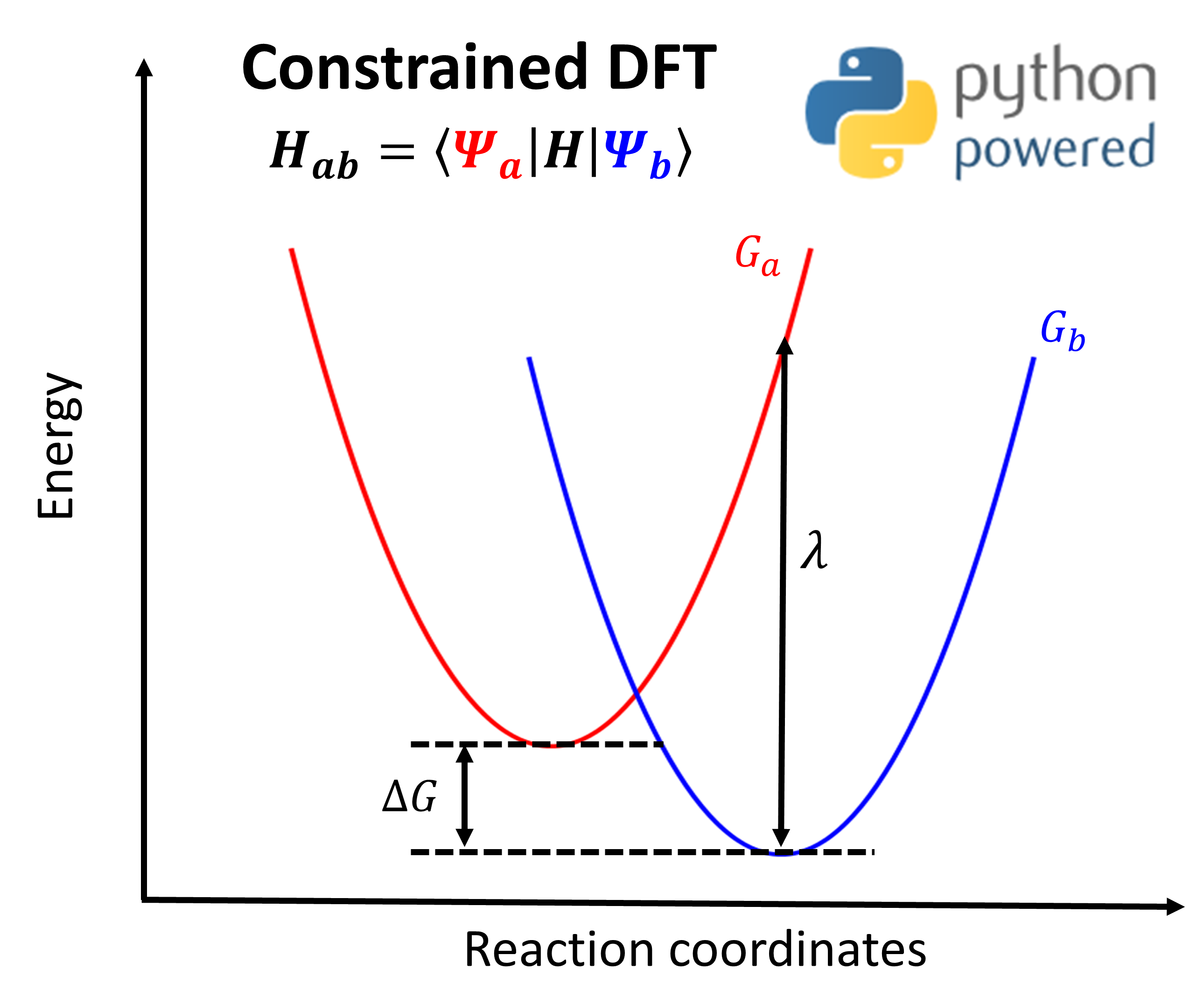} 
\\
Charge transfer plays a critical role in many physical, chemical, and biological processes, and can be described using the coupling between diabatic states.
In this work, we present \texttt{PyCDFT}, an open-source Python package that is robust, flexible, and DFT-engine agnostic for calculating diabatic states and their electronic coupling using constrained density functional theory.
\end{minipage}
}}
\end{figure}

  \makeatletter
  \renewcommand\@biblabel[1]{#1.}
  \makeatother

\bibliographystyle{apsrev}

\renewcommand{\baselinestretch}{1.5}
\normalsize

\clearpage

\section{\sffamily \Large INTRODUCTION} 
\label{sec:intro}

The transfer of electronic charges plays a central role in many physical and chemical processes~\cite{mayChargeAndEnergy2011}, such as those for cellular activity in biological processes~\cite{blumbergerRecentAdvancesTheory2015} and catalytic activity in condensed phases~\cite{newtonElectronTransferReactions1984}. In addition,  the rate of charge transfer in a material directly impacts its carrier mobility and hence its use in e.g., electronic devices~\cite{jiangUnderstandingCarrierTransport2019,jiangNegativeIsotopeEffect2015}.

Theoretical and computational modeling  provides invaluable insights into the microscopic mechanism of charge transfer, and is playing an important role in the development of novel drugs, catalysts, and electronic materials. In the past few decades, many research efforts have been dedicated to the development of robust theoretical methods and simulation strategies to describe charge transfer processes in molecules and materials~\cite{hopfieldElectronTransferBiological1974,jonesNatureTunnelingPathway2002,zengFragmentBasedQuantumMechanical2012,gilletElectronicCouplingCalculations2016,yamadaEffectiveTightBindingModels2010,schoberCriticalAnalysisFragmentorbital2016,oberhoferChargeTransportMolecular2017}.
Charge transfer can take place through a wide spectrum of mechanisms, with two important regimes being the band-like regime (where transport occurs through delocalized electronic states) and the hopping regime (where transport occurs through localized electronic states)~\cite{shuaiTheoryOfCharge2012,Bässler2012}.
Here we focus on the hopping transfer, which is the dominant charge transfer mechanism in many organic crystals and conducting polymers, and in several metal oxides in the solid state, as well as in many nanoparticle solids~\cite{lanQuantumDotSolids2020,talapinProspectsColloidalNanocrystals2010,guyot-sionnestElectricalTransportColloidal2012,yuVariableRangeHopping2004}.

The classic theory of charge transfer in the hopping regime is Marcus theory~\cite{marcusTheoryOxidationReduction1956,marcusElectronTransferReactions1993}, which has seen many generalizations through the years~\cite{landauTheoryTransferEnergy1932,zenerNonAdiabaticCrossingEnergy1932, linUltrafastDynamics2002,nanNuclearTunnelingEffects2009}.
For a charge transfer between two sites A and B (e.g., a donor-acceptor pair consisting of two molecules or two  fragments of the same molecular unit), Marcus theory predicts the charge transfer rate to be
\begin{equation}
    k = \frac{2\pi}{\hbar} |H_{ab}|^2 \sqrt{\frac{1}{4k_BT\pi\lambda}} \exp \left[ - \frac{(\Delta G + \lambda)^2}{ 4\lambda k_B T} \right],
\end{equation}
where the diabatic electronic coupling $H_{ab}$ between A and B is one of the central quantities that determines transfer rates; $k_B$ and T are the Boltzmann constant and temperature; $\Delta G$ is the free energy difference between states A and B, and $\lambda$ is the reorganization energy.
As shown in Fig.~\ref{fig:cc-et}, within Marcus theory the charge transfer process can be described using the free energy surfaces of two \textit{diabatic states} as functions of a chosen reaction coordinate.
Diabatic states are defined as a set of states among which the nonadiabatic derivative couplings vanish.
Diabatic states have the property that their physical characters (such as charge localization) do not change along the reaction coordinate.
For instance, the two diabatic states ($\Psi_a$/$\Psi_b$) involved in the charge transfer depicted in Fig. \ref{fig:cc-et} are constructed to have the charge localized on site A/B, and this charge localization character does not change as the reaction occurs.


In contrast to \textit{adiabatic} states, which are the eigenstates of the electronic Hamiltonian within the Born-Oppenheimer approximation, \textit{diabatic} states are \textit{not} eigenstates of the electronic Hamiltonian of the whole system, and therefore are not directly accessible from standard electronic structure calculations.
Constrained density functional theory (CDFT) provides a powerful and robust framework for constructing diabatic states from first principles and predicting their electronic coupling~\cite{kadukConstrainedDensityFunctional2012, goldeyChargeTransportNanostructured2017}, including instances where hybrid functionals may fail to produce a localized state~\cite{melanderImplementationConstrainedDFT2016} and where time-dependent DFT may fail to produce the correct spatial decay of the electronic coupling~\cite{dreuwFailureTimeDependentDensity2004}.
In CDFT, additional constraint potentials are added to the Kohn-Sham Hamiltonian, and their strengths are optimized so as to obtain a desired localized charge on a given site.
To obtain the electronic coupling $H_{ab}$, one first performs two separate CDFT calculations in which one localizes the charge on the initial and final sites.
Then, one constructs the electronic Hamiltonian matrix on the basis composed of the two diabatic states, and finally the $H_{ab}$ is given by the off-diagonal elements of the Hamiltonian matrix.

A CDFT formulation was originally proposed by Dederichs in 1984~\cite{dederichsGroundStatesConstrained1984} to study excitations of Ce impurities in metals.
Wu, van Voorhis and co-workers established the modern formulation of CDFT in the mid-2000s~\cite{wuExtractingElectronTransfer2006a,wuDirectCalculationElectron2006}.
Since then it has been implemented in several DFT codes using localized basis sets, such as \textsc{Siesta}~\cite{souzaConstrainedDFTMethodAccurate2013}, NWChem~\cite{wuDirectCalculationElectron2006}, Q-CHEM~\cite{wuConstrainedDensityFunctional2009} and ADF~\cite{ADF2001}.

Implementations of CDFT using plane-wave basis sets appeared more recently, for instance in CPMD~\cite{oberhoferChargeConstrainedDensity2009,oberhoferElectronicCouplingMatrix2010}, VASP~\cite{maConstrainedDensityFunctional2015} and CP2K (dual basis)~\cite{hutterCP2KAtomisticSimulations2014}.
These plane-wave implementations enabled CDFT calculations for condensed systems, and facilitated the study of important problems such as redox couples in aqueous solution~\cite{blumbergerDiabaticFreeEnergy2006,oberhoferChargeConstrainedDensity2009,oberhoferChargeConstrainedDensity2009}, charge transfer in biological molecules and proteins~\cite{oberhoferRevisitingElectronicCouplings2012}, in quantum dots~\cite{brawandDefectStatesCharge2017} and doped nanoparticles~\cite{vorosHydrogenTreatmentDetergent2017}, electron tunneling between defects~\cite{blumbergerConstrainedDensityFunctional2013} and polaron transport~\cite{seoRolePointDefects2018,wangRoleSurfaceOxygen2020} in oxides, molecular solids~\cite{oberhoferRevisitingElectronicCouplings2012}, and organic photovoltaic polymers~\cite{goldeyPlanarityMultipleComponents2016} (see Ref.~\citenum{kadukConstrainedDensityFunctional2012} and  Ref.~\citenum{blumbergerRecentAdvancesTheory2015} for extensive reviews). In existing implementations, DFT and CDFT are developed and maintained in the same code, thus requiring  direct modifications of core DFT routines to support CDFT functionalities.

In recent years, an emerging trend in scientific simulation software is the development of light-weight code, with focus on specific tasks, which can be interfaced with other codes to perform complex tasks.
This strategy is well aligned with the modular programming coding practice, which enables maintainability, re-usability, and simplicity of codes.
Compared to conventional strategies integrating a wide range of functionalities into one single code, this design strategy decouples the development cycle of different functionalities and leads to inter-operable codes that are easier to modify and maintain, facilitating rapid developments and release of new features.
Some notable codes for chemical and materials simulations that have adopted this strategy include Qbox~\cite{gygiArchitectureQboxScalable2008}, WEST~\cite{govoniLargeScaleGW2015,maFiniteFieldApproach2018}, and SSAGES~\cite{sidkySSAGES2018,sevgenHierarchicalCouplingFirstPrinciples2018}.

In this work we present \texttt{PyCDFT}, a Python package that performs single-point self-consistent-field (SCF) and geometry optimization calculations using CDFT. \texttt{PyCDFT} can be interfaced with existing DFT codes to perform DFT calculations with constraint potentials. Compared to existing implementations of CDFT, the novelty of the \texttt{PyCDFT} code is twofold:

\begin{itemize}
\item \texttt{PyCDFT} is a light-weight, interoperable code.
The operations specific to CDFT calculations are decoupled from those carried out by existing DFT codes (DFT engines).
Communications between \texttt{PyCDFT} and the DFT engine are handled by client-server interfaces (see Sec.~\ref{sec:software}).
Hence, the development of \texttt{PyCDFT} and of the DFT engine may occur independently.
This is advantageous for maintainability and reusability, and \texttt{PyCDFT} may be interfaced with multiple DFT engines.

\item \texttt{PyCDFT} features an object-oriented design that is user-friendly and extensible. Extra functionalities can be easily added to \texttt{PyCDFT} thanks to the extensive use of abstract classes. Furthermore, \texttt{PyCDFT} supports being used within Jupyter notebooks or Python terminals, thus allowing users to perform and analyze CDFT calculations in a flexible and interactive manner.

\end{itemize}

\noindent We note that Python has become increasingly popular as a high-level programming language for scientific computing due to its ease of use and wide applicability. The development of \texttt{PyCDFT} echos this trend and contributes to the rapidly expanding open-source Python ecosystem for the molecular and materials science fields, where some widely-used packages include Atomic Simulation Environment (\texttt{ASE})~\cite{hjorthlarsenAtomicSimulationEnvironment2017}, \texttt{pymatgen}~\cite{ongPythonMaterialsGenomics2013}, and \texttt{PySCF}~\cite{sunPYSCF2018}.


To demonstrate the use of \texttt{PyCDFT}, we coupled it with the massively parallel first-principles molecular dynamics code \texttt{Qbox}~\cite{gygiArchitectureQboxScalable2008}, which features efficient DFT calculations using plane-wave basis sets and pseudopotentials.
We computed diabatic electronic coupling for a set of organic molecules in the HAB18 data set~\cite{kubasElectronicCouplingsMolecular2014,kubasElectronicCouplingsMolecular2015} and compared our results with those of existing implementations.
The results obtained with \texttt{PyCDFT}(Qbox) are in good agreement with those of other plane-wave implementations of CDFT, thus verifying the correctness and robustness of \texttt{PyCDFT}.

\section{\sffamily \Large COMPUTATIONAL METHODOLOGY}
\subsection{\sffamily \large Constrained Density Functional Theory}
\label{sec:theory}

We briefly outline the CDFT methodology adopted here and we refer the reader to Refs.~\citenum{kadukConstrainedDensityFunctional2012, oberhoferElectronicCouplingMatrix2010, wuDirectOptimizationMethod2005,wuDirectCalculationElectron2006,melanderImplementationConstrainedDFT2016} for further details.
The core of the CDFT method is the iterative calculation of the stationary point of a free energy functional $W$ defined as
\begin{equation} \label{eq:W}
    W[n, V_k] = E[n] + \sum_k V_k \left( \int w_k(\mathbf{r}) n(\mathbf{r}) d\mathbf{r} - N_k^0 \right),
\end{equation}
where $n$ is the electron density; $E[n]$ is the DFT total energy functional; the second term on the right-hand side of Eq.~\ref{eq:W} represents the sum of constraint potentials applied to the system to ensure that the desired number of electrons $N_k^0$ is localized on given parts of the system (e.g., chosen atomic site, molecule, or structural fragment).
More than one constraint can be applied to the system, if needed.
The strength of the $k^\mathrm{th}$ constraint potential is controlled by the scalar Lagrange multiplier $V_k$, and its shape is determined by a weight function $w_k(\mathbf{r})$.
CDFT calculations are performed by self-consistently minimizing $W$ with respect to $n$ and maximizing $W$ with respect to $V_k$.
The minimization of $W$ with respect to $n$ is equivalent to performing a DFT calculation with additional constraint potentials $\sum_k V_k w_k(\mathbf{r})$ added to the Kohn-Sham Hamiltonian.
Upon convergence of the SCF cycle, the number of electrons localized on a given site $N_k = \int d\mathbf{r} \, w_k(\mathbf{r}) n(\mathbf{r})$ is equal to the desired value $N_k^0$.
In geometry optimization calculations, the free energy $W$ is further minimized with respect to nuclear coordinates, as shown in the outermost cycle in Fig.~\ref{fig:workflow}.

\subsection{\sffamily \large Calculation of weight functions}
\label{sec:hirsh}

The weight function allows one to partition the total electron density into contributions from different fragments of the whole system.
Several different partitioning schemes have been proposed, such as Mulliken ~\cite{mullikenElectronicPopulationAnalysis1955}, Becke ~\cite{beckeMulticenterNumericalIntegration1988}, and Hirshfeld partitioning~\cite{hirshfeldBondedatomFragmentsDescribing1977}.
In \texttt{PyCDFT} we implemented the Hirshfeld partitioning, which is widely used in plane-wave implementations of CDFT~\cite{kubasElectronicCouplingsMolecular2014,kubasElectronicCouplingsMolecular2015,goldeyChargeTransportNanostructured2017}.
The Hirshfeld weight function $w$ is defined as the ratio between the pseudoatomic densities belonging to a given site and the total pseudoatomic density
\begin{equation} \label{eq:hirshfeld}
    w(\boldsymbol{r}) = \frac{\sum_{I \in F} \rho_I(\mathbf{r} - \mathbf{R}_I)}{\sum_I \rho_I(\mathbf{r} - \mathbf{R}_I)},
\end{equation}
where $I$ denotes atoms and $I \in F$ denotes atoms belonging to a fragment $F$ to which the constraint is applied; $\mathbf{R}_I$ is the coordinate of atom $I$; $\rho_I$ denotes the electron density of the  \textit{isolated} $I$-th atom and should not be confused with the electron density $n$ of the whole system.

Alternatively, to enforce constraints on the electron number difference between a donor site D and an acceptor site A, one can define the weight function as:
\begin{equation} \label{eq:hirshfeld_ct}
    w(\mathbf{r})= \frac{\sum_{I \in D} \rho_I(\mathbf{r} - \mathbf{R}_I)-\sum_{I \in A} \rho_I(\mathbf{r} - \mathbf{R}_I)}{\sum_I \rho_I(\mathbf{r} - \mathbf{R}_I)}.
\end{equation}

Both definitions of Hirshfeld weights are implemented in \texttt{PyCDFT} [see Sec. S3 of the supplementary information (SI)]. 
For charge transfer processes where the whole system consists of only two fragments (donor and acceptor), the above two definitions of Hirshfeld weights are equivalent.
For more complex processes where multiple parts of the system are involved, one can use a combination of the two definitions to enforce complex charge constraints.

In Eqs. \ref{eq:hirshfeld} and Eq. \ref{eq:hirshfeld_ct}, the real-space electron density of an atom located at $R_I$ is computed as:
\begin{equation}
    \rho_I(\mathbf{r} - \mathbf{R}_I) =  4\pi \mathcal{F}^{-1} \left[ e^{-i\mathbf{G} \cdot \mathbf{R}_I} \int_0^\infty \rho_I(r) \frac{r \sin(Gr)}{G} dr \right],
\end{equation}
where $\mathcal{F}^{-1}$ denotes an inverse Fourier transform; $\mathbf{G}$ is  a reciprocal lattice vector with norm $G$; $\rho_I(r)$ is the radial electron density of atom $I$.
For a given atomic species, $\rho_I(r)$ can be easily obtained by performing DFT calculations for isolated atoms.
\texttt{PyCDFT} is distributed with pre-computed spherically-averaged electron densities obtained with the SG15 pseudopotentials~\cite{hamannOptimizedNormconservingVanderbilt2013,schlipfOptimizationAlgorithmGeneration2015} for all species in the periodic table before bismuth (excluding the lanthanides).

\subsection{\sffamily \large Calculation of Forces}
\label{sec:forces}

In order to perform geometry optimizations or molecular dynamics simulations on a  diabatic potential energy surface, the force on each nucleus due to the applied constraints must be evaluated.
Forces on the diabatic potential energy surface are the sum of  the  DFT forces $F^\text{DFT}$ and the constraint force $F^\text{c}$ arising from the derivative of the constraint potential with respect to nuclear coordinates.

For a system subject to constraints, the $\alpha$ component ($\alpha \in \{x, y, z\}$) of the constraint force $F^\text{c}$ on the $I^\text{th}$ atom is given by:

\begin{equation}\label{eq:forces}
\begin{split}
    F^{\text{c}}_{I\alpha} &= - \sum_k V_k \int d\mathbf{r} \, \rho(\mathbf{r}) \frac{\partial w_k(\mathbf{r})}{\partial R_{I\alpha}}  \\
  & = - \sum_k V_k \int d\mathbf{r} \, \rho(\mathbf{r}) \frac{\delta - w_k(\mathbf{r})}{\sum_J \rho_J (\mathbf{r} - \mathbf{R}_J)} \frac{\partial \rho_I (\mathbf{r} - \mathbf{R}_I)}{\partial R_{I\alpha}},
\end{split}
\end{equation}
where $\delta = \delta_{I \in F}$ for constraints on absolute electron numbers (Eq. \ref{eq:hirshfeld}) and $\delta = \delta_{I \in D} - \delta_{I \in A}$ for constraints on electron number differences (Eq. \ref{eq:hirshfeld_ct}).
The term $\frac{\partial \rho_I (\mathbf{r} - \mathbf{R}_I)}{\partial R_{I\alpha}}$ is evaluated as:
\begin{equation}\label{eq:drhodr}
\frac{\partial \rho_I (\mathbf{r} - \mathbf{R}_I)}{\partial R_{I\alpha}} = \mathcal{F}^{-1} \left\{ - i G_\alpha e^{-i \mathbf{G} \cdot \mathbf{R}_I} \mathcal{F} \left[ \rho_I(\mathbf{r}) \right] \right\},
\end{equation}
where $\mathcal{F}$ and $\mathcal{F}^{-1}$ denote forward and backward Fourier transforms, respectively.
In Sec. S3 of the SI, we verify the analytical calculation of forces using Eq.~\ref{eq:forces} and Eq.~\ref{eq:drhodr} by comparing with results obtained with finite difference calculations of total energies.

\subsection{\sffamily \large Diabatic electronic coupling}
\label{sec:coupling}

To compute the electronic coupling $H_{ab} $\cite{wuExtractingElectronTransfer2006a}, we consider the Hamiltonian matrix on the diabatic basis composed of two diabatic states $\Psi_a$ and $\Psi_b$, each obtained from a converged CDFT calculation with \texttt{PyCDFT}.
Here we consider the case of a single constraint.
Denoting the value of the Lagrange multiplier for the two CDFT calculations as $V_a$ and $V_b$, respectively, the Hamiltonian on the diabatic basis is:
\begin{equation}
  \bf{H} = \left(\begin{matrix}  H_{aa} & H_{ab} \\ H_{ba} & H_{bb} \end{matrix}\right),
\end{equation}
where the diagonal elements $H_{aa}$ and $H_{bb}$ correspond to the DFT total energies of diabatic states $\Psi_a$ and $\Psi_b$, respectively. Then, denoting the overlap matrix $\mathbf{S}$ between the two diabatic states as
\begin{equation}
  \mathbf{S} = \left(\begin{matrix}  1 & S_{ab} \\ S_{ba} & 1 \end{matrix}\right),
\end{equation}
where $S_{ab} = \langle \Psi_a | \Psi_b \rangle $ and $S_{ab} = S_{ba}^*$, the off-diagonal Hamiltonian elements can be written as~\cite{oberhoferElectronicCouplingMatrix2010}:
\begin{equation}
  H_{ab} = F_b S_{ab} - V_b W_{ab}
\end{equation}
\begin{equation}
  H_{ba} = F_a S_{ba} - V_a W_{ba}
\end{equation}
where $F_a$ and $F_b$ are the CDFT total energies including the contribution of constraint potentials; the weight function matrix elements $W_{ab}=W_{ba}^*$ are given by $W_{ab} = \langle \Psi_a |w(\mathbf{r})| \Psi_b \rangle$.

After $\mathbf{H}$ is evaluated in the diabatic basis, we follow Ref.~\citenum{oberhoferElectronicCouplingMatrix2010} and average the off-diagonal elements of $\mathbf{H}$ to ensure its Hermiticity.
Finally, we perform a L\"owdin orthogonalization~\cite{lowdinNonOrthogonalityProblem1950} for $\mathbf{H}$ using the overlap matrix $\mathbf{S}$
\begin{equation}
  \mathbf{\tilde{H}} = \mathbf{S}^{-1/2} \mathbf{H} \mathbf{S}^{-1/2}
\end{equation}
and the off-diagonal matrix element of $\mathbf{\tilde{H}}$ corresponds to the electronic coupling $H_{ab}$.

\section{\sffamily \Large Software}
\label{sec:software}

\subsection*{\sffamily \large Implementation}

\texttt{PyCDFT} features an object-oriented design and extensive use of abstract classes and abstract methods to facilitate future extensions of functionalities.
Here we list the major classes defined in the \texttt{PyCDFT} package.

\begin{itemize}
    \item \texttt{Sample}: a container class to organize relevant information about the physical system. A \texttt{Sample} instance is constructed by specifying the positions of the atoms within the periodic cell. The \texttt{Sample} class utilizes the \texttt{ASE}~\cite{hjorthlarsenAtomicSimulationEnvironment2017} package to parse atomic structures from geometry files (e.g., \texttt{cif} files).

    \item \texttt{Fragment}: a container class to represent a part of the whole system to which constraints are applied. A \texttt{Fragment} instance is constructed by specifying a list of atoms belonging to the fragment.

    \item \texttt{Constraint}: an abstract class representing a constraint applied to the system. A \texttt{Constraint} instance keeps track of physical quantities relevant to the constraint, such as $N^0_k$, $N_k$, $V_k$, and $w_k(\mathbf{r})$ (see Eq.~\ref{eq:W}). Except for the parameter $N^0_k$, which is defined upon the construction of the instance, other quantities are updated self-consistently as the CDFT calculation proceeds. Currently, two types of constraints based on Hirshfeld partitioning are implemented: \texttt{ChargeConstraint} (Eq.~\ref{eq:hirshfeld}) and \texttt{ChargeTransferConstraint} (Eq.~\ref{eq:hirshfeld_ct}). 

    \item \texttt{DFTDriver}: an abstract class that controls how \texttt{PyCDFT} interacts with an external DFT code. It specifies how \texttt{PyCDFT} communicates the constraint potentials and constraint forces to the DFT code and how to fetch the charge densities and other relevant quantities from the DFT code. Currently, a subclass \texttt{QboxDriver} is implemented, which allows \texttt{PyCDFT} to interact with the \texttt{Qbox} code. The implementation of the \texttt{QboxDriver} class leverages the client-server interface of \texttt{Qbox}, which allows \texttt{Qbox} to interactively respond to commands provided by a user or an external code~\cite{maFiniteField2018,nguyenFiniteField2019} (\texttt{PyCDFT} in this case).

    \item \texttt{CDFTSolver}: the core class of \texttt{PyCDFT} that executes a CDFT calculations. \texttt{CDFTSolver} provides a \texttt{solve} method, which is used to perform a CDFT self-consistent or geometry optimization calculation. Optimization of the Lagrange multipliers is performed within the \texttt{solve} method, which utilizes the \texttt{scipy} package.

\end{itemize}

In addition to the above classes, \texttt{PyCDFT} contains a \texttt{compute\_elcoupling} function, which takes two \texttt{CDFTSolver} instances as input and computes the electronic coupling $H_{ab}$ between  two diabatic states (see Sec.~\ref{sec:coupling}).
To enable the calculation of electronic coupling, \texttt{PyCDFT} implements an auxiliary \texttt{Wavefunction} class that stores and manipulates the Kohn-Sham orbitals from CDFT calculations.

 \subsection*{\sffamily \large Extensibility}

Thanks to the use of abstract classes, \texttt{PyCDFT} can be easily extended to provide new functionalities.
For instance, support for additional weight functions (such as spin-dependent weight functions) can be easily implemented by defining subclasses of \texttt{Constraint} and overriding its abstract methods.
Similarly, one can extend \texttt{PyCDFT} to support other DFT codes by overriding the abstract methods in the \texttt{DFTDriver} class.
In addition to the C++ code \texttt{Qbox} used here, several Python implementations of DFT (e.g., \texttt{PySCF}) may be called as a DFT driver in an interactive manner; therefore they may be used as DFT drivers of \texttt{PyCDFT} once the corresponding \texttt{DFTDriver} subclass is implemented.

\texttt{PyCDFT} may also be readily integrated with existing Python-based interfaces for generating, executing, and analyzing electronic structure calculations using software such as \texttt{ASE}~\cite{hjorthlarsenAtomicSimulationEnvironment2017} and \texttt{Atomate}~\cite{mathewAtomateHighlevelInterface2017}.

\subsection{\sffamily \large Installation and usage}
\label{sec:install}

Installation of \texttt{PyCDFT} follows the standard procedure using the \texttt{setup.py} file included in the distribution.
Currently, it depends on a few readily available Python packages including \texttt{ASE}, \texttt{scipy}, \texttt{pyFFTW}, and \texttt{lxml}.

In Fig.~\ref{fig:code} we present an example script that utilizes \texttt{PyCDFT} to compute the diabatic electronic coupling for the helium dimer $\mathrm{He}_2^+$.
This and other examples are included in the distribution of \texttt{PyCDFT}.


\section{\sffamily \Large Verification}
\label{sec:verification}

We now turn to the verification of our implementation of CDFT in \texttt{PyCDFT}, focusing on the calculation of electronic couplings. We compare results obtained with  \texttt{PyCDFT}(Qbox), CPMD~\cite{kubasElectronicCouplingsMolecular2014,kubasElectronicCouplingsMolecular2015,oberhoferElectronicCouplingMatrix2010}), CP2K, and the implementation of CDFT in \textsc{Quantum ESPRESSO}\cite{giannozziAdvancedCapabilitiesMaterials2017} originally contributed by Goldey et al.~\cite{goldeyChargeTransportNanostructured2017}. We note that all codes utilized for this comparison use plane-wave basis sets, with the exception of CP2K, which uses a mixed Gaussian and plane-wave basis set.
As the values obtained for the electronic coupling have been shown to be sensitive to the choice of weight partitioning schemes~\cite{oberhoferElectronicCouplingMatrix2010}, we compare with only results obtained with the Hirshfeld partitioning scheme.

Our results, \texttt{PyCDFT}(Qbox), are obtained by performing DFT calculations with the \texttt{Qbox}~\cite{gygiArchitectureQboxScalable2008} code. We used optimized norm-conserving Vanderbilt pseudopotentials (ONCV)~\cite{hamannOptimizedNormconservingVanderbilt2013,schlipfOptimizationAlgorithmGeneration2015}, and an energy cutoff of 40 Ry for all molecules; we tested up to a 90 Ry energy cutoff and found changes of 1-2\% in the electronic coupling compared to calculations using a 40 Ry cutoff. We used a convergence threshold of $5 \times 10^{-5}$ for $|N - N_0|$.
The electronic couplings were converged to within less than 0.5\% with respect to cell size, in order to minimize interactions with periodic images.
When using  CP2K, we adopted the TZV2P basis set with GTH pseudopotentials~\cite{goedeckerSeparableDualSpace1996}. Results obtained with {\sc Quantum Espresso} (QE) and CPMD have been previously reported in Ref.~\citenum{goldeyChargeTransportNanostructured2017} and Refs.~\citenum{kubasElectronicCouplingsMolecular2014,kubasElectronicCouplingsMolecular2015}, respectively.
In all cases the DFT electronic structure problem was solved using the generalized gradient approximation of Perdew, Burke, and Ernzerhof (PBE)~\cite{perdewGeneralizedGradientApproximation1996}.

We first discuss results for the electronic coupling of the He$_2^+$ dimer. 
Figure~\ref{fig:he-dimer} compares the decay in $H_{ab}$ with distance for hole transfer in the He-He$^+$ dimer obtained with \texttt{PyCDFT}(Qbox) and other codes.
We find excellent agreement between our computed electronic couplings and those from Oberhofer and Blumberger~\cite{oberhoferElectronicCouplingMatrix2010} obtained using CPMD and the results of Goldey et al.~\cite{goldeyChargeTransportNanostructured2017} obtained using QE.
As wavefunctions decay exponentially, the variation of the electronic coupling with separation may be expressed as $H \propto \exp(-\beta R/2)$, and we can compare the decay behaviors obtained here and in the literature by using the decay rate $\beta$, which is found to be 4.64, 4.98, 4.13 1/\AA\, with
 \texttt{PyCDFT}(Qbox), CPMD, and {\sc Quantum Espresso} (QE), respectively.



We now turn to bench-marking results for molecular dimers in the HAB18 dataset, which combines the HAB11~\cite{kubasElectronicCouplingsMolecular2014} and HAB7~\cite{kubasElectronicCouplingsMolecular2015} data sets, and consists of $\pi$-stacked organic homo-dimers.
The molecules in the HAB11 data set contain members with different number of $\pi$-bonds and atomic species; the HAB7 dataset contains larger molecules.
The combined HAB18 data set has been previously used for other implementations of CDFT~\cite{goldeyChargeTransportNanostructured2017}.
The first molecule we consider here is one where  imperfect $\pi$-stacking is present, due to one of the monomers being rotated relative to the other.
Fig.~\ref{fig:thio-dimer} compares our calculated electron coupling for this configuration of the thiophene dimer with that of Kubas et al as implemented in CPMD~\cite{kubasElectronicCouplingsMolecular2014}.
We find excellent agreement between the two results, thus demonstrating the accuracy and robustness of \texttt{PyCDFT}(Qbox) for off-symmetry configurations.

Next we analyze in greater detail the performance of \texttt{PyCDFT}(Qbox) compared to CP2K, CPMD, and Quantum Espresso implementations for other members of the HAB18 data set for perfectly $\pi$-stacked homo-dimers.
In the appendix, Table S1 (see SI) shows the calculated electronic couplings for molecular dimers as a function of inter-molecular distance. 
Tables S2 and S3 report the mean error, mean absolute error, root-mean-square deviation, and mean absolute percent error among codes  for the electronic coupling and decay constants, respectively. 


We compare our computed electronic couplings of molecular dimers in the HAB18 data set at varying intermolecular distances using \texttt{PyCDFT}(Qbox) with those obtained with CP2K, CPMD, and QE.
These are plotted in Fig.~\ref{fig:habplot} on a log-log scale.
In general, there is good agreement among the various codes.
There is a systematic deviation of all DFT results from those based on multi-reference configuration interaction (MRCI+Q)\cite{kubasElectronicCouplingsMolecular2014} and single-determinant spin-component-scaled coupled cluster (SCS-CC2)\cite{kubasElectronicCouplingsMolecular2015} calculations.
This systematic deviation arises from the well-known delocalization error of the semi-local functional used here (PBE) and from its shortcoming to properly describe long-range dispersion interactions.
Using more accurate functionals would improve the accuracy of CDFT, as previously reported in the literature~\cite{kubasElectronicCouplingsMolecular2015}.
Nevertheless, inspection of Fig.~\ref{fig:habplot} (and Table S1) shows that \texttt{PyCDFT}(Qbox) generally yields electronic couplings and decay constants within the range of values obtained from previous implementations. 
Finally, we emphasize that \texttt{PyCDFT}(Qbox) captures the physically relevant exponential decay of the electronic coupling with intermolecular distance.

\section{\sffamily \Large CONCLUSIONS}

In this work we presented \texttt{PyCDFT}, a Python module for performing calculations based on constrained density function theory (CDFT).
\texttt{PyCDFT} allows for SCF and geometry optimization calculations of diabatic states, as well as calculations of diabatic electronic couplings.
The implementation of CDFT in \texttt{PyCDFT} is flexible and modular, and enables ease of use, maintenance, and effective dissemination of the code.
Using molecules from the HAB18 data set~\cite{kubasElectronicCouplingsMolecular2014,kubasElectronicCouplingsMolecular2015} as benchmarks, we demonstrated that \texttt{PyCDFT}(Qbox) yields results in good agreement with those of existing CDFT implementations using  plane-wave basis sets and pseudopotentials.
As a robust implementation for CDFT calculations, \texttt{PyCDFT} is well-suited for first-principles studies of charge transfer processes.

\subsection*{\sffamily \large ACKNOWLEDGMENTS}

We thank Francois Gygi for helpful discussions. We thank Chenghan Li for generous help with calculations using \texttt{CP2K}.
H.M., M.G. and G.G. are supported by MICCoM, as part of the Computational Materials Sciences Program funded by the U.S. Department of Energy, Office of Science, Basic Energy Sciences, Materials Sciences and Engineering Division through Argonne National Laboratory, under contract number DE-AC02-06CH11357. W.W. is supported by the National Science Foundation (NSF) under Grant No. CHE-1764399. S.K. was supported by DOE/BES under under grant no. DE-SC0012405.
This research used computational resources of the University of Chicago Research Computing Center.

\clearpage

\bibliography{library}


\clearpage

\begin{figure}
\caption{\label{fig:cc-et} Free energy curves for two diabatic states $\Psi_a$ and $\Psi_b$ with free energy $G_a$ and $G_b$ associated to a reaction where a charge (electron or hole) is transferred from site A to site B. The charge is localized on site A for $\Psi_a$ and site B for $\Psi_b$, and the charge localization characters of $\Psi_a$ and $\Psi_b$ do not change as the reaction occurs. The charge transfer rate can be written as a function of the free energy difference $\Delta G$, reorganization energy $\lambda$, and electronic coupling $H_{ab}$ (see text). }
\end{figure}

\begin{figure}
\caption{\label{fig:workflow} Workflow for self-consistent-field (SCF) and geometry optimization calculations performed by \texttt{PyCDFT}. In SCF calculations, the free energy functional $W$ is minimized with respect to the electron density $n$ (equivalent to a standard DFT calculation under constraint potentials) and maximized with respect to Lagrange multipliers $V_k$. For geometry optimization calculations, $W$ is further minimized with respect to nuclear coordinates $\mathbf{R}$. \texttt{PyCDFT} is designed to implement CDFT-specific algorithms and to be interfaced with external DFT codes (drivers).}
\end{figure}

\begin{figure}
  \caption{\label{fig:code} An example Python script to perform CDFT calculations for $\mathrm{He}_2^+$. Two \texttt{CDFTSolver} instances are created for the calculation of two diabatic states with different charge localization, then the \texttt{compute\_elcoupling} function is called to compute the electronic coupling $H_{ab}$ between the two diabatic states.}
\end{figure}

\begin{figure}
     \caption{\label{fig:he-dimer} Comparison of diabatic electronic coupling $H_{ab}$ of the He-He+ dimer as a function of distance $R$, calculated with constrained density functional theory, and using PyCDFT interfaced with the Qbox code (\texttt{PyCDFT}(Qbox)), the implementation of CDFT in CPMD from Oberhofer and Blumberger~\cite{oberhoferElectronicCouplingMatrix2010}, and the implementation in {\sc Quantum Espresso} (QE) from Goldey et al~\cite{goldeyChargeTransportNanostructured2017}. In all implementations, the Hirshfeld partitioning~\cite{hirshfeldBondedatomFragmentsDescribing1977} scheme is used. The calculated $\beta$ decay rates are 4.64, 4.98, and 4.13 1/\AA\, respectively.}
\end{figure}

\begin{figure}
    \caption{\label{fig:thio-dimer} Diabatic electronic coupling $H_{ab}$ of the stacked thiophene dimer at a separation of 5 \AA\,  as a function of the relative rotation of the two units, calculated with constrained density functional theory as implemented in this work (\texttt{PyCDFT}(Qbox)) and in Kubas et al in CPMD~\cite{kubasElectronicCouplingsMolecular2014}. Carbon atoms are shown in brown, sulfur in yellow, and hydrogen in beige.}
\end{figure}

\begin{figure}
       \caption{\label{fig:habplot} Log-log plot of computed diabatic electronic couplings for molecular dimers in the HAB18 data set~\cite{kubasElectronicCouplingsMolecular2014,kubasElectronicCouplingsMolecular2015} at various inter-molecular distances using \texttt{PyCDFT}(Qbox) (blue circles), CP2K (purple stars), CPMD (green squares), and Quantum Espresso (QE, yellow triangles). Reference values (black line) are based on multi-reference configuration interaction (MRCI+Q)\cite{kubasElectronicCouplingsMolecular2014} and single-determinant spin-component-scaled coupled cluster (SCS-CC2)\cite{kubasElectronicCouplingsMolecular2015} level of theory.}
\end{figure}


\clearpage

\vspace*{0.1in}   
\begin{center}
\includegraphics[width=3.5in]{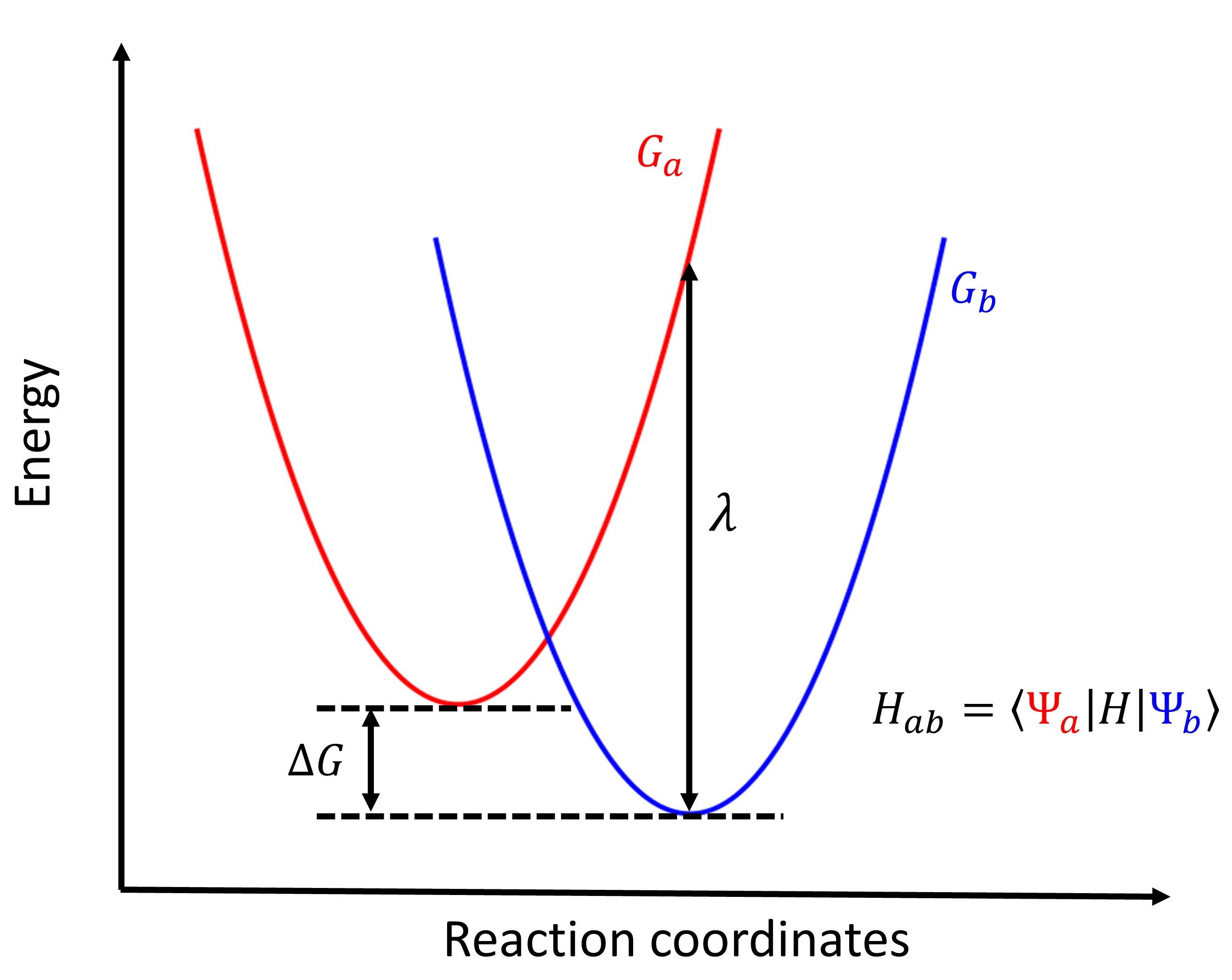}
\end{center}
\vspace{0.25in}
\hspace*{3in}
{\Large
\begin{minipage}[t]{3in}
\baselineskip = .5\baselineskip
Figure 1 \\
He Ma, Wennie Wang, Siyoung Kim, Man-Hin Cheng, Marco Govoni, Giulia Galli \\
J.\ Comput.\ Chem.
\end{minipage}
}

\newpage
\vspace*{0.1in}   
\begin{center}
    \includegraphics[width=5in]{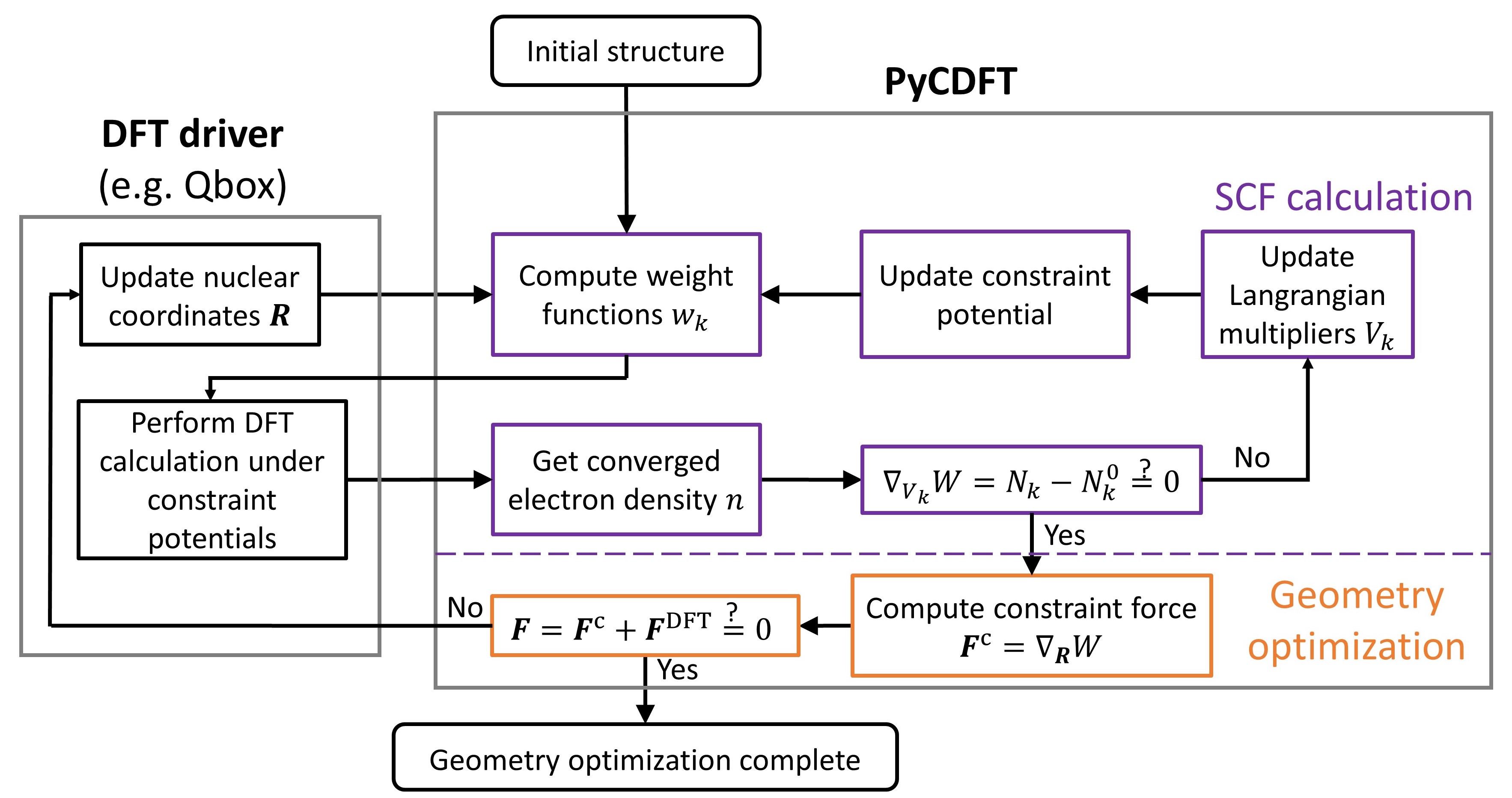}
\end{center}
\vspace{0.25in}
\hspace*{3in}
{\Large
\begin{minipage}[t]{3in}
\baselineskip = .5\baselineskip
Figure 2 \\
He Ma, Wennie Wang, Siyoung Kim, Man-Hin Cheng, Marco Govoni, Giulia Galli \\
J.\ Comput.\ Chem.
\end{minipage}
}

\newpage
\vspace*{0.1in}   
\begin{center}
  \includegraphics[width=6in]{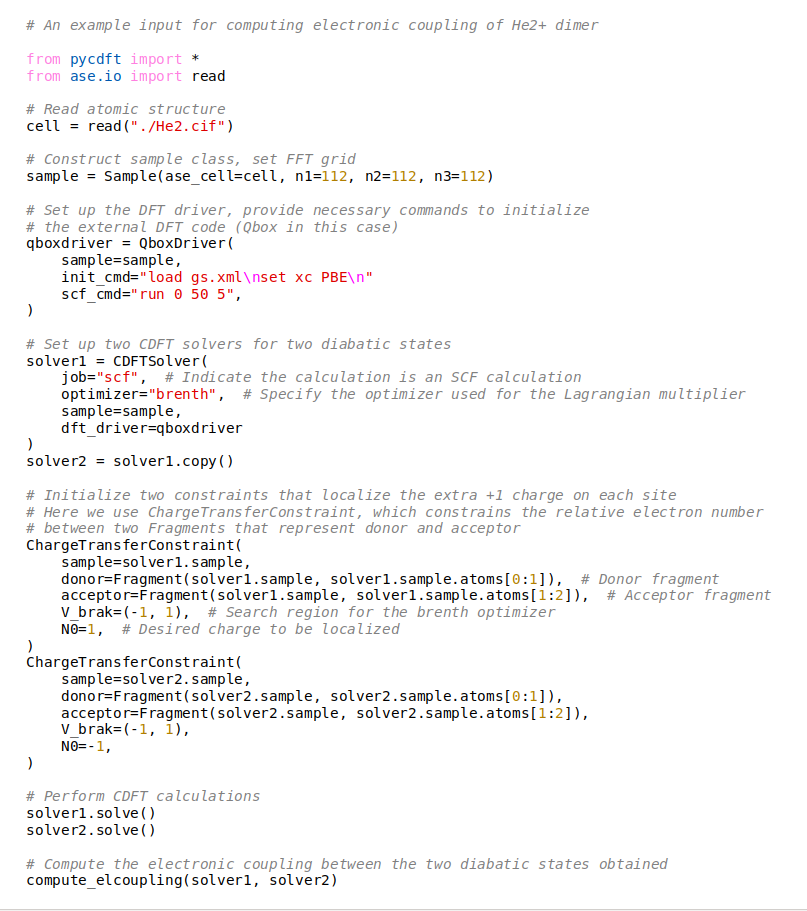}
\end{center}
\vspace{0.25in}
\hspace*{3in}
{\Large
\begin{minipage}[t]{3in}
\baselineskip = .5\baselineskip
Figure 3 \\
He Ma, Wennie Wang, Siyoung Kim, Man-Hin Cheng, Marco Govoni, Giulia Galli \\
J.\ Comput.\ Chem.
\end{minipage}
}

\newpage
\vspace*{0.1in}   
\begin{center}
  \includegraphics[width=0.6\columnwidth]{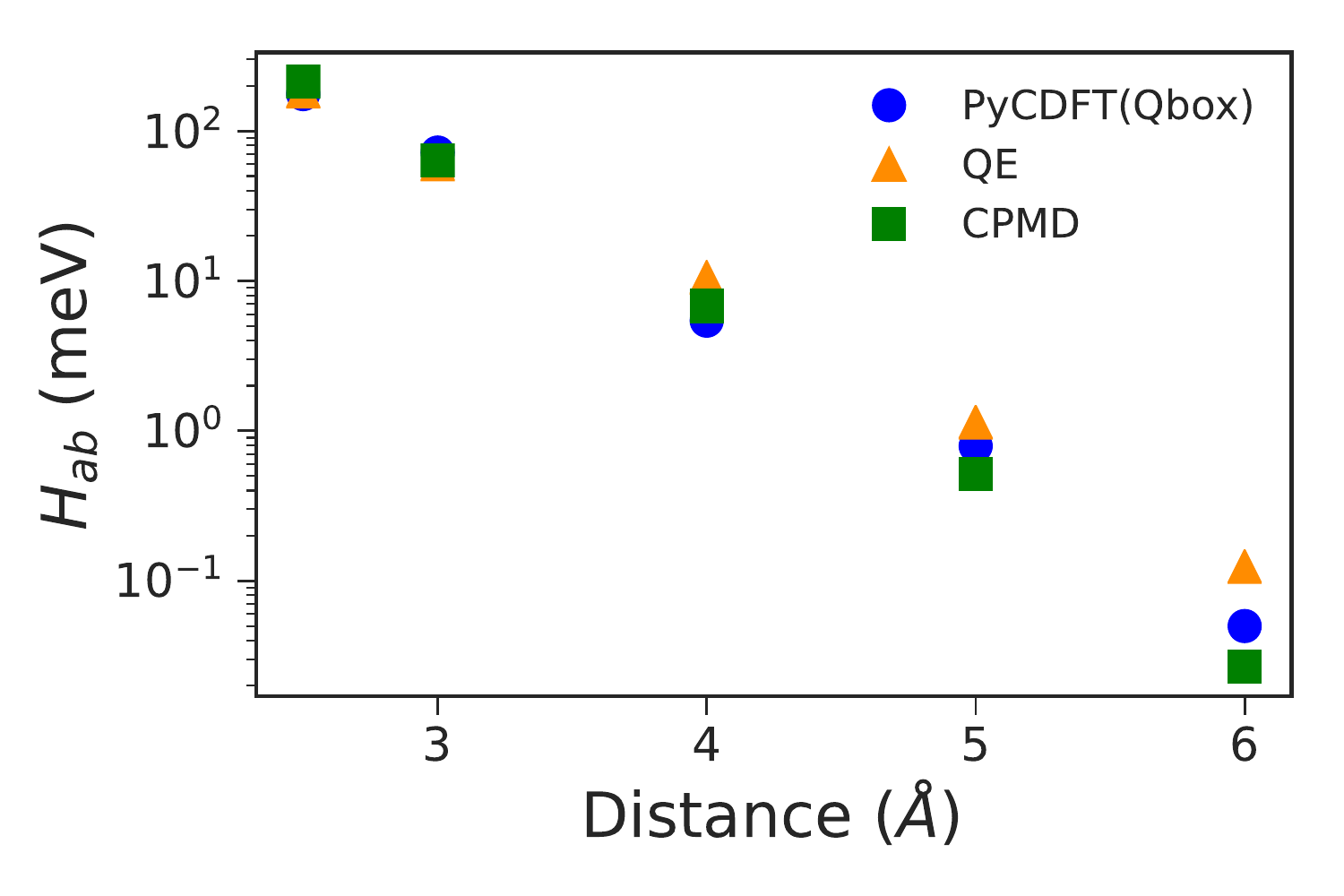}
\end{center}
\vspace{0.25in}
\hspace*{3in}
{\Large
\begin{minipage}[t]{3in}
\baselineskip = .5\baselineskip
Figure 4 \\
He Ma, Wennie Wang, Siyoung Kim, Man-Hin Cheng, Marco Govoni, Giulia Galli \\
J.\ Comput.\ Chem.
\end{minipage}
}

\vspace*{0.1in}   
\begin{center}
  \includegraphics[width=0.6\columnwidth]{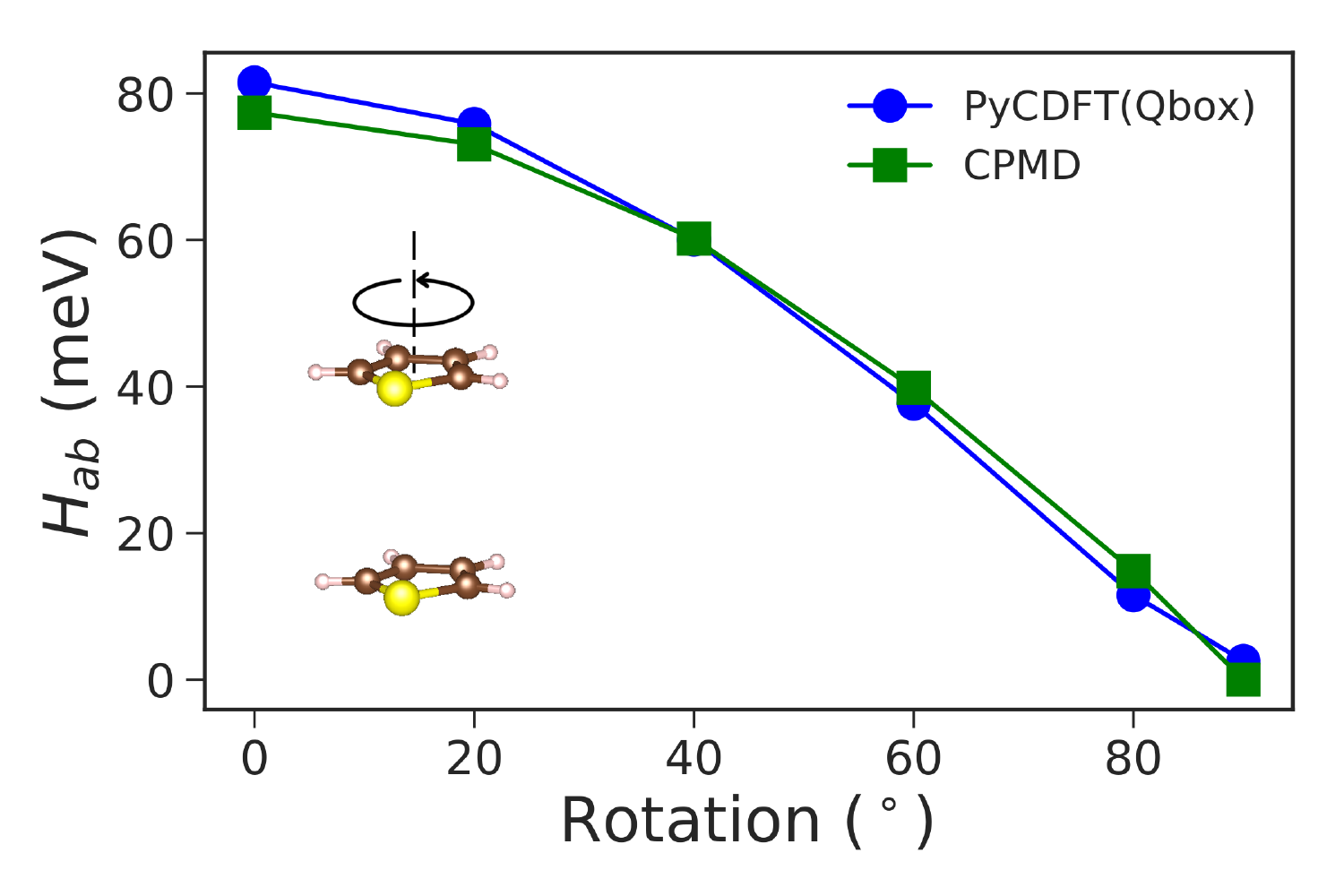}
\end{center}
\vspace{0.25in}
\hspace*{3in}
{\Large
\begin{minipage}[t]{3in}
\baselineskip = .5\baselineskip
Figure 5 \\
He Ma, Wennie Wang, Siyoung Kim, Man-Hin Cheng, Marco Govoni, Giulia Galli \\
J.\ Comput.\ Chem.
\end{minipage}
}

\vspace*{0.1in}   
\begin{center}
  \includegraphics[width=4in]{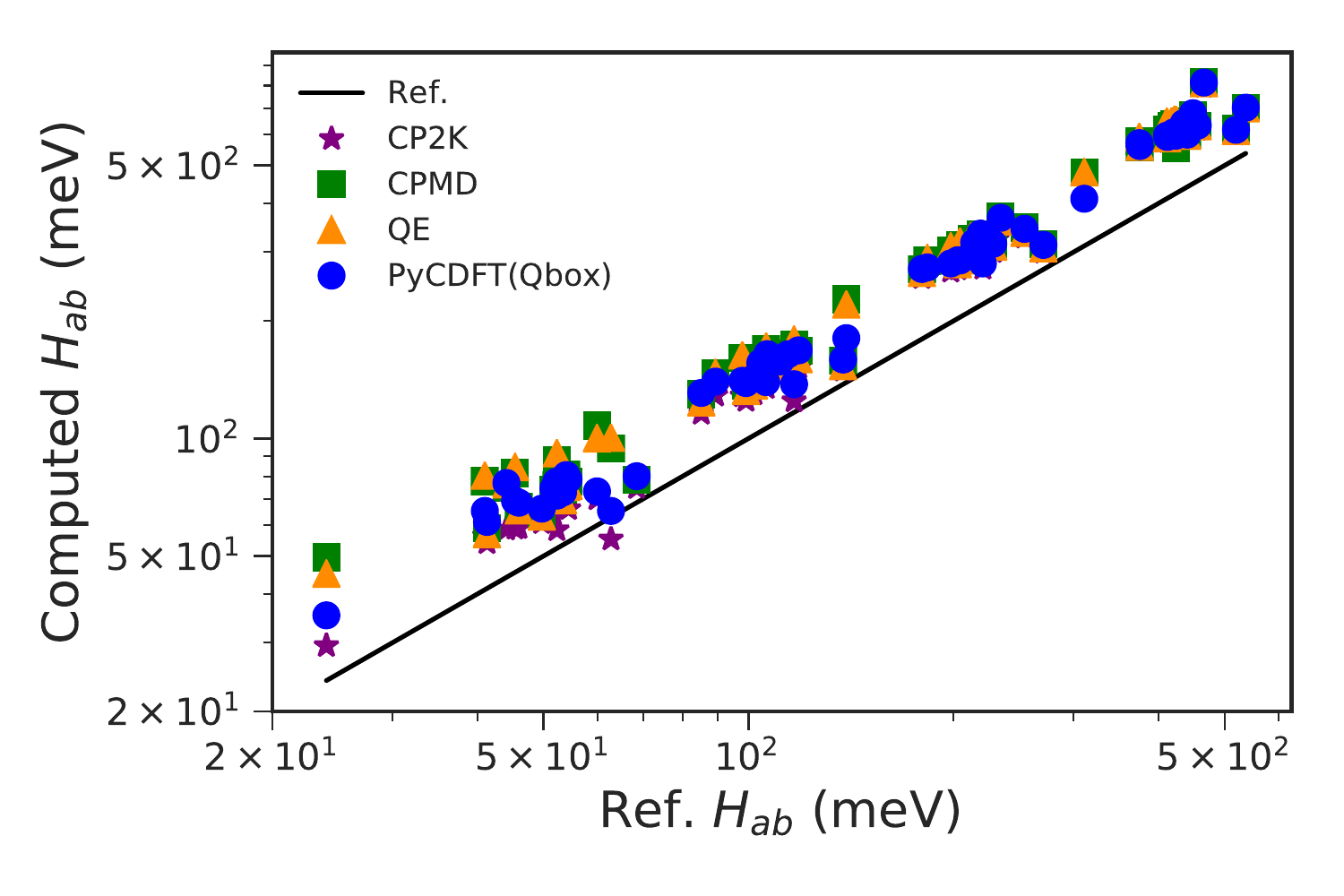}
\end{center}
\vspace{0.25in}
\hspace*{3in}
{\Large
\begin{minipage}[t]{3in}
\baselineskip = .5\baselineskip
Figure 6 \\
He Ma, Wennie Wang, Siyoung Kim, Man-Hin Cheng, Marco Govoni, Giulia Galli \\
J.\ Comput.\ Chem.
\end{minipage}
}

\end{document}